\documentclass[twocolumn,floatfix,pre,aps,showpacs]{revtex4}
\input{graphicx}
\def \ind#1{{\mbox{\footnotesize {#1}}}}
\def \ind#1{{\mbox{\scriptsize {#1}}}}

\begin{document}

\title{Heterogeneities in the Glassy State}
\author{Katharina Vollmayr-Lee$^{1,2}$ and Annette Zippelius$^{2}$}
\email{kvollmay@bucknell.edu}
\affiliation{$^{1}$ Department of Physics, Bucknell University,  
      Lewisburg, Pennsylvania 17837, USA\\
   $^{2}$ Institut f\"ur Theoretische Physik, Universit\"at G\"ottingen,
       Friedrich-Hund-Platz 1, D-37077 G\"ottingen, Germany}
\date{July 5, 2005}

\begin{abstract}
  We study heterogeneities in a binary Lennard-Jones system
  {\em below} the glass transition using molecular dynamics
  simulations.  We identify mobile and immobile particles and measure
  their distribution of vibrational amplitudes.  For temperatures near
  the glass transition the distribution of vibrational amplitudes
  obeys scaling and compares reasonably well with a mean field theory
  for the amorphous solid state. 
  To investigate correlations among the immobile and mobile particles
  we identify clusters and analyze their size and shape.  For a fixed
  number of immobile particles we observe 
  that the immobile particles cluster more strongly together
  as the temperature is increased
  which allows the particles to block each other more 
  effectively, and to therefore stay immobile.
  For the mobile particles, on the other hand,
  the clustering is most pronounced at small temperatures, indicating
  that mobility at low temperatures can only be sustained in 
  cooperative motion.
\end{abstract}

\maketitle

\section{Introduction}
\label{sec:intro} 
If a liquid is cooled and crystallization is avoided one obtains 
a supercooled liquid. Upon further cooling the system falls out of
equilibrium and results in an amorphous solid, i.e.\ a glass. During this
transition from liquid to supercooled liquid to glass the dynamic
properties change drastically and they therefore have received much 
attention. Most studies are concerned with the supercooled liquid, for which
the existence of dynamic heterogeneities has been well established 
\cite{ref:dynhet_reviews}. We focus in this paper instead on the {\em glass}
phase.

The glassy state of matter is by definition strongly inhomogeneous.
Different particles have different neighborhoods and consequently
different dynamics. One tool to diagnose the heterogeneities are the
vibrational amplitudes $\langle d_i^2 \rangle$. 
In contrast to a perfect crystal these vary
strongly from site to site in the glassy state. A simple way to
demonstrate the heterogeneities is to pick out the $5\%$ particles
which have the smallest and largest amplitudes. Their vibrational
amplitudes are plotted in Fig.~\ref{fig:disq_perc} for a binary
Lennard-Jones mixture (for details see below). The most striking 
observation is the spread in the data which strongly increases with increasing
temperature. Whereas for $T=0.15$ the $\langle d_i^2 \rangle$ of the 5\%
fastest particles  is approximately five times the $\langle d_i^2
\rangle$ of the 5\% slowest particles, this grows to a factor of 
approximately fifty 
close to the glass transition temperature.
Thus, heterogeneities are present in the glassy state for all
temperatures and become increasingly more pronounced as the glass
transition temperature is approached from below.

Our goal is to characterize these heterogeneities quantitatively.
Obviously the mean of the vibrational amplitudes is not conclusive and
we therefore investigate instead the full distribution of vibrational
amplitudes to characterize the local dynamics. We compute this
distribution and show that, in the vicinity of the glass transition,
it approximately obeys scaling. We furthermore compare it to a mean
field calculation for a different amorphous solid
\cite{ref:Pofdisq_MF}, the gel, and find surprisingly good agreement.

To further characterize dynamic as well as time persistent 
heterogeneities we use various 
definitions to identify mobile and immobile particles.
We identify clusters of immobile and mobile particles 
where particles are defined to be
connected  if their relative distance is smaller than
the position of the first minimum of the pair correlation function.
To quantify the shape of
these clusters we analyze their connectivity properties in
analogy to covalently bonded glasses. We determine the average size
of the clusters, the average coordination number and the spatial extent
as measured by the radius of gyration.  In addition to the average
values we also compute the distribution of coordination numbers as
well as ring statistics.
 
The majority of previous work on dynamic heterogeneities are studies of
{\em mobile} particles {\em above} the glass transition. They find that
mobile particles form string-like clusters 
\cite{ref:Donati1998,ref:PRE60,ref:JCP111,ref:JCP120,ref:Teboul} 
which grow in length and size with decreasing
temperature \cite{ref:mob_length+size,ref:PRE70,ref:eweeks}.
Immobile particles on the other hand do not form strings
\cite{ref:PRE60,ref:JCP111}. In this paper we mostly study clusters
of {\em immobile} particles {\em below} the glass transition and quantify 
their size and shape as described above.

\begin{figure}[tbp]
\includegraphics[width=3.2in]{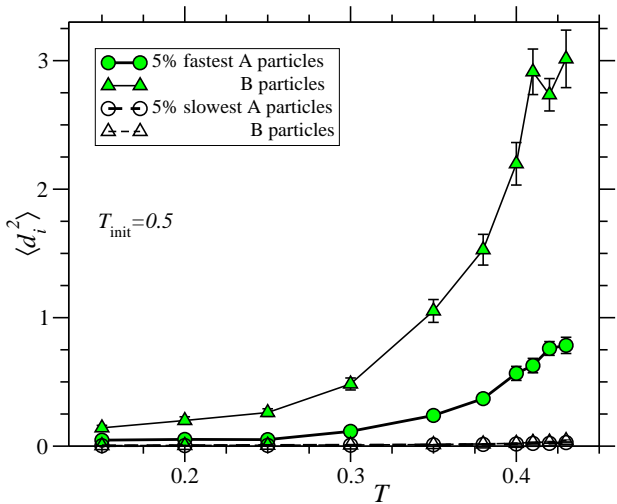}
\caption{\sf Vibrational amplitude $\langle d_i^2 \rangle$ as a function 
  of temperature for the 5\% fastest and 5\% slowest particles. 
See Fig.~\protect\ref{fig:disq_cut+nojump} for 
an enlargement of $\langle d_i^2 \rangle (T)$ for the 5\% slowest 
A and B particles.
}
\label{fig:disq_perc}
\end{figure}

The paper is organized as follows. In Sec.~\ref{sec:simulation} we give
details of the simulation. Subsequently several different definitions
of immobile particles are introduced and discussed
(Sec.~\ref{sec:def_immobile}). In Sec.~\ref{sec:results} we present our
results, first for the fraction of immobile particles (A), next for
the vibrational amplitudes (B) and finally the cluster analysis (C).
We summarize our results and present conclusions in
Sec.~\ref{sec:conclusions}.

\section{Simulation Details}
\label{sec:simulation}

Our system is a binary Lennard-Jones mixture of 800 A and 200 B 
particles with the same mass. The interaction potential 
between particles $i$ and $j$ of type $\alpha, \beta \in \{$A,B$\}$ 
and at positions ${\bf r}_i$ and ${\bf r}_j$ is
\begin{equation}  
\label{eq:Valphabeta}
V_{\alpha \beta}(r) = 4 \, \epsilon_{\alpha \beta} \,
\left ( \left ( \frac{\sigma_{\alpha \beta}}{r} \right )^{12}
      - \left ( \frac{\sigma_{\alpha \beta}}{r} \right )^{6}
\right ),
\end{equation}
where $r = |{\bf r}_i - {\bf r}_j|$ and
$\epsilon_\ind{AA}=1.0$, $\epsilon_\ind{AB}=1.5$, $\epsilon_\ind{BB}=0.5$, 
$\sigma_\ind{AA}=1.0$, $\sigma_\ind{AB}=0.8$ and
$\sigma_\ind{BB}=0.88$. We truncate and shift the potential at
$r=2.5 \,\sigma_{\alpha \beta}$. From previous investigations
\cite{ref:kob_andersen} it is known
that this system is not prone to crystallization and demixing.
In the following we
will use reduced units where the unit of length is $\sigma_\ind{AA}$,
the unit of energy is $\epsilon_\ind{AA}$ and the unit of time is
$\sqrt{m \sigma_\ind{AA}^2/(48 \epsilon_\ind{AA})}$.

We carry out molecular dynamics (MD) simulations using the velocity
Verlet algorithm with a time step of 0.02. The volume is kept constant
at $V=9.4^3=831$ and we use periodic boundary conditions.  Previous
simulations \cite{ref:kob_andersen} showed that
computer simulations the system falls out of
equilibrium in the vicinity of $T_{\ind{g}} = 0.435$. We are interested in the
glassy phase and hence study the system well below and close to
this temperature in the range $0.15 \leq T \leq 0.43$.

As described in Sec.~IX of \cite{ref:JCP2004} we equilibrate the system 
at constant temperature $T=3.0$ to obtain ten independent configurations
which are at least $5 \cdot 10^4$ time units apart. The temperature is
kept constant by replacing the velocities of all particles by new
velocities drawn from the corresponding Boltzmann distribution
every 50 time steps. Each of these configurations is then cooled
linearly in time $t$ ($T(t)=T_0 - \gamma \,t$) from $T_0=3.0$
to $T_\ind{init}=0.5$ with $\gamma=1.25 \cdot 10^{-5}$. After an (NVT)
equilibration at $T=0.5$ of $2 \cdot 10^5$ time units, we then
instantaneously quench the system to the investigated temperatures 
$T=0.15, 0.20, 0.25, 0.30, 0.35, 0.38, 0.40, 0.41, 0.42$ and 
$0.43$. We anneal the system at each temperature with (NVT) simulations
for 2000 time units and then run the production runs with (NVE)
simulations for $10^5$ time units. Since the relaxation times $\tau$ 
are much larger than the waiting time before the production runs 
($\tau \approx 8\cdot 10^5$ at $T=0.446$), we expect to find aging
effects as they have been studied in detail for the same Lennard-Jones
system \cite{ref:Kob4581,ref:Barrat637} and as they have been found 
in related work 
\cite{ref:JCP2004,ref:JCP2002}. 
In this paper we do not aim for a systematic investigation of aging
effects. However, we have analyzed a second set of simulation runs
with a different history in order to get a rough estimate of the
effects of aging.  The second set of runs start from well equilibrated
configurations at $T_\ind{init}=0.446$. After an instantaneous quench
to $T=0.15$ the system is annealed for 2000 time units followed by a
(NVT) run of also 2000 time units. The temperature is then raised to
$T=0.2$ again annealed with (NVT) and (NVE) runs each of 2000 time
units and raised to $T=0.25$ etc.. The production runs are for $10^5$
time units and are for each investigated temperature following the
respective (NVE) simulations. For more details see
\cite{ref:JCP2004,ref:JCP2002}

\section{Definition of Immobile and Mobile Particles}
\label{sec:def_immobile}

We study heterogeneities below the glass transition and
focus on how the particles are increasingly frozen in as the
temperature is lowered. Even though the system is obviously very
inhomogeneous (see Sec.~\ref{sec:intro}), 
it is not obvious how to quantify the
characteristic features of these
heterogeneities. One possibility is to study the extremes, the most
mobile and the most immobile particles. In most of this paper we
concentrate on the immobile ones; we study their organization into
clusters and the size and shape of these clusters.

Using four different definitions of immobility we are able 
to test if the resulting dynamical behavior is dependent on 
the precise definition of immobility. For two of the definitions
we characterize the mobility of a particle by its vibrational amplitude
\cite{ref:JCP2002}
\begin{equation} \label{eq:disq} 
d_i^2 = \overline{ \left |{\bf r}_i - \overline{{\bf r}_i} \right |^2 }
\end{equation}
where the bar denotes a time average over the whole simulation run.
Our first definition of immobility (I) defines every particle $i$ to be
immobile for which 
\begin{equation} \label{eq:defI}
d_i^2 \le d^2_\ind{cut}
\end{equation}
The second definition of immobility (II) \cite{ref:JCP2002} 
identifies as immobile particles a fixed percentage of 
A particles and separately a fixed percentage of  B particles with the smallest
$d_i^2$. This allows us to similarly study the 
most mobile particles with the largest $d_i^2$. 

\begin{figure}[tbp]
\includegraphics[width=3.2in]{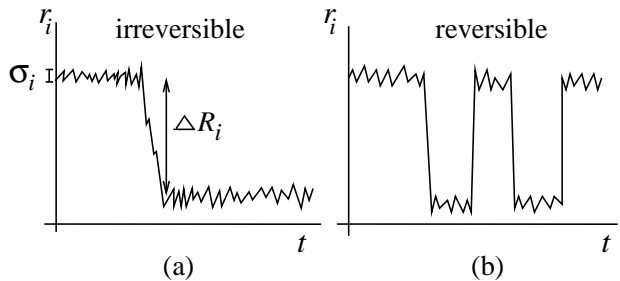}
\caption{\sf  Sketch of a particle trajectory to illustrate the 
  definition of irreversible and reversible jumps.
}
\label{fig:jump_def_2}
\end{figure}
The third and fourth definition of immobility are based on recent work 
on single particle motion \cite{ref:JCP2004}. As sketched 
in Fig.~\ref{fig:jump_def_2}a, we define a particle $i$ to undergo a jump
if its change in average position $\Delta R_i$ is significantly larger
than its fluctuation $\sigma_i$ ($\Delta R_i > \sqrt{20} \, \sigma_i$). We
distinguish irreversible and reversible jumps where in the latter case
the particle returns during the simulation run to one of its former
average positions (for more details see \cite{ref:JCP2004}). We call all
particles immobile which are not undergoing any jumps (III)
or which are not irreversible jumpers (IV).

Note that our definitions of immobile particles refer to the whole
simulation run, because we want to investigate those structures which
persist on the longest timescale, set by the length of the
simulation run.

\section{Results}
\label{sec:results}

\subsection{Fraction of Immobile Particles}
\label{sec:fraction_immobile}

\begin{figure}[tbp]
\includegraphics[width=3.2in]{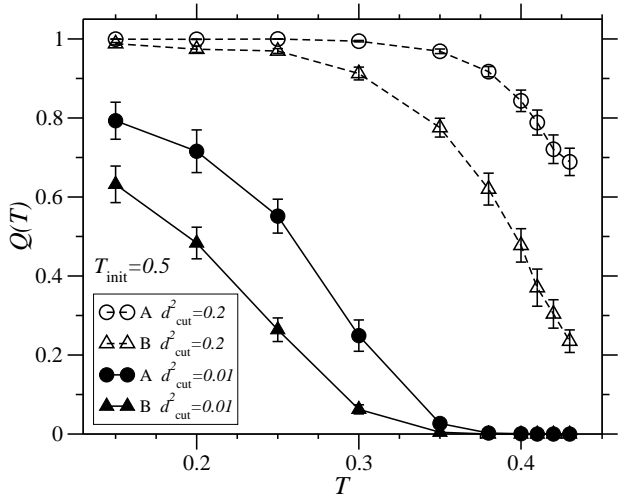}
\caption{\sf The fraction of immobile particles $Q$ as a function of 
temperature $T$ using definition I for immobility. 
$Q(T)$ is given both for 
A and B particles separately and for $d^2_\ind{cut}=0.2$ and 
$d^2_\ind{cut}=0.01$.
The simulations are with initial Temperature $T_\ind{init}=0.5$
}
\label{fig:QofT_dcut}
\end{figure}
\begin{figure}[tbp]
\includegraphics[width=3.2in]{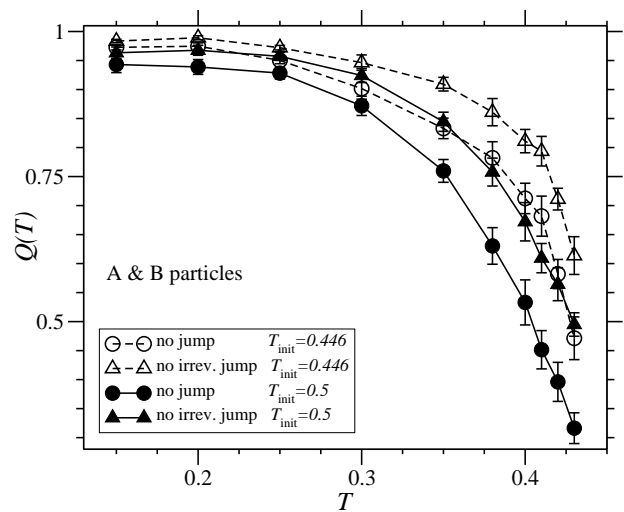}
\caption{\sf Same as Fig.~\ref{fig:QofT_dcut} but for definitions III 
  and IV, for A and B particles, and for different initial temperatures.
}
\label{fig:QofT_nojump}
\end{figure}
Let us first see how many particles are immobile according to the
definitions I, III and IV. Fig.~\ref{fig:QofT_dcut} and
\ref{fig:QofT_nojump} show that the fraction of immobile particles 
\begin{equation}
\label{eq:Q}
Q = \mbox{number of immobile part. / number of part.}
\end{equation}
approaches one for very low temperatures and drops significantly for
temperatures near the glass transition.  $Q(T)$ depends quantitatively
on the definition of immobile: the larger $d^2_\ind{cut}$ the steeper
is the drop of $Q$.  The smallest cutoff $d^2_\ind{cut}=0.01$
corresponds to the Lindemann criterion and gives rise to significant
deviations in the results for $Q(T)$ as compared to the other
definitions of immobility. Apparently the frozen structure can be
sustained with a considerable fraction of particles whose vibrational
amplitude is larger than suggested by the Lindemann criterion, e.g.
roughly $25\%$ of the A-particles at $T=0.3$.  This is consistent with
the broad scatter of vibrational amplitudes in the glassy phase that
we shall discuss below. 
 
The qualitative behavior is the same for all definitions of immobility
and in particular the temperature where $Q$ goes to zero is rather
insensitive to the details of the definition of immobile particles
(unless the cutoff is chosen too small, see above). Hence
$Q$ serves as a good indicator for $T_\ind{g}$.
Similarly the history of the simulation runs has a quantitative
influence on $Q(T)$: larger initial temperature ($T_\ind{init}=0.5$)
results in fewer immobile particles, consistent with Fig.~22 of
\cite{ref:JCP2004}. The qualitative behavior, however, is not only
independent of the precise definition of immobility but also
independent of the history of the simulations runs.

\subsection{Vibrational Amplitude}
\label{sec:disq}

\begin{figure}[tbp]
\includegraphics[width=3.2in]{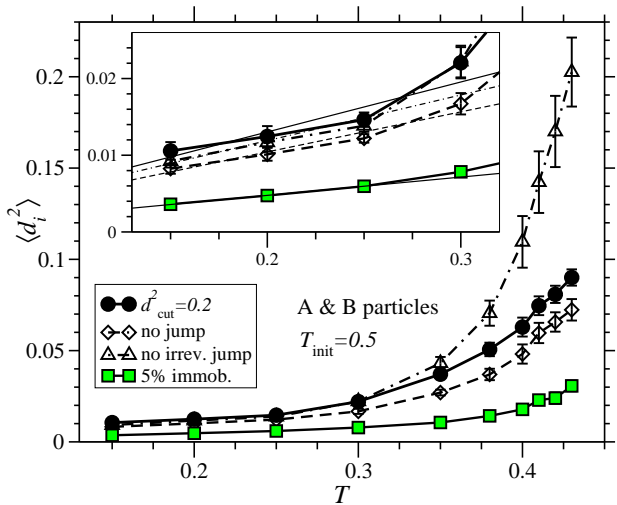}
\caption{\sf Same as Fig.~\ref{fig:disq_perc} for immobile 
  particles using definitions 
  I - IV and for A and B particles. The inset shows an 
enlargement for low temperatures with
 linear fits through the origin.
}
\label{fig:disq_cut+nojump}
\end{figure}

One possibility to extract the various length scales of the glassy
state are the vibrational amplitudes $\{d_i^2\}$. 
Fig.~\ref{fig:disq_perc} shows $\langle d_i^2 \rangle$ as a function of
temperature $T$ for the 5\% particles with largest $d_i^2$ (mobile,
filled symbols) 
and the 5\% particles with smallest $d_i^2$ (immobile, open symbols).  
Here $\langle \cdot \rangle$
denotes an average over 
the subset of particles under consideration (mobile or immobile)
and over initial configurations \cite{ref:errorbars}.
The remarkable spread of $\langle d_i^2 \rangle$ in Fig.~\ref{fig:disq_perc}
illustrates the increasing importance  of 
heterogeneities when the glass transition temperature is approached from
below (see Sec.~\ref{sec:intro}).
In the inset of Fig.~\ref{fig:disq_cut+nojump} we have included linear 
fits through the origin, which would
correspond to the dynamics of a harmonic solid.
We find 
for the 5\% immobile particles $\langle d_i^2 \rangle (T)$ for $T \le 0.25$ 
good agreement with this line fit.
For larger temperatures, however, the immobile particles display 
anharmonic dynamics. 
On the other hand, the immobile particles of definitions I, III, and IV 
include at low temperatures almost all particles (see 
Fig.~\ref{fig:QofT_dcut} and Fig.~\ref{fig:QofT_nojump}), and 
the linear fits through the origin are therefore only good approximations 
for $T \le 0.2$.

The $\langle d_i^2 \rangle$ of immobile particles using definitions I,
III and IV are slightly larger but comparable to the $\langle d_i^2
\rangle$ using definition II (see Fig.~\ref{fig:disq_cut+nojump}). We
find for all definitions of immobility the same temperature dependence.
Similarly the analogous figures for $T_\ind{init}=0.446$ are
qualitatively the same. 

\begin{figure}[tbp]
\includegraphics[width=3.2in]{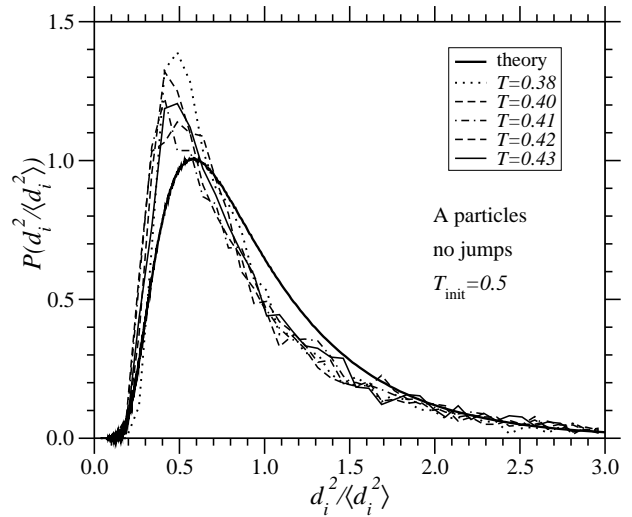}
\caption{\sf Distribution of the normalized vibrational amplitude 
  compared to a mean field theory \cite{ref:Pofdisq_SH}. For the simulation 
 data the definition III for immobility has been used and the 
  distribution is shown for A particles.
}
\label{fig:Pofdisqovdisqav_nojump}
\end{figure}

To investigate further the heterogeneities we show in
Fig.~\ref{fig:Pofdisqovdisqav_nojump} the distribution of the
vibrational amplitude normalized by its average. The distribution of
vibrational amplitudes has been computed analytically for the
amorphous solid which results from crosslinking macromolecules, i.e.\ 
the gel state \cite{ref:Pofdisq_SH,ref:Pofdisq_MF}. 
In that context $Q$ corresponds
to the gel fraction and $\langle d_i^2 \rangle$ to the localization
length squared.  It was shown that the distribution of localization
length obeys scaling in the critical regime of the transition from the
fluid to the amorphous solid state which is a second order equilibrium
phase transition.  Furthermore, the scaling function has been computed
within a mean field approach. The analytical results are in good
agreement with results from numerical simulations of crosslinked
macromolecular melts \cite{ref:simul_gels}.

The two systems, binary Lennard-Jones glasses and crosslinked
polymeric melts, have completely different microscopic dynamics and
the nature of the glass transition, if it exists, is not understood in
contrast to the gelation transition. Nevertheless it is an interesting
open question to see whether the distribution of localization lengths 
$P(d_i^2)$ obeys scaling close to the glass transition and how it
compares to the universal function of the gelation transition.
Fig.~\ref{fig:Pofdisqovdisqav_nojump} illustrates that 
indeed for temperatures $T \ge 0.38$ $P(d_i^2)$ depends on
temperature only through the average squared localization length and
that the normalized distribution compares reasonably well with the universal
function of the gelation transition, as predicted by mean field theory 
\cite{ref:Pofdisq_SH,ref:Pofdisq_MF}.
This agreement with theory is independent of the
definition of immobility and independent of the history of initial
configurations.

\subsection{Cluster Analysis}
\label{sec:cluster}

The vibrational amplitude has given us an estimate for the
length scale of single particle motion. In the rest of the paper
we investigate how the immobile and mobile particles are spatially
correlated with each other and how they are organized in clusters. 

To do so we use definitions I -- IV of immobility (see Sec.~III) to
select a subset of particles of the system. We stress again that
immobility defined in this way is a time persistent feature (on the
timescale of the simulation run). We then choose a particular instant
of time --- the beginning of each production run --- to identify nearest
neighbor connections among this subset of particles: particle $j$ is
defined to be a neighbor of particle $i$ if their distance $|{\bf
  r}_{ij}| = |{\bf r}_i - {\bf r}_j|$ is smaller than the position of
the first minimum $r_\ind{min}$ of the corresponding radial pair
distribution function of the complete system ($r_\ind{min}=1.4$ for
AA, $1.2$ for AB and $1.07$ for BB independent of temperature).

This definition of nearest neighbor connections gives rise to $K$
distinct clusters of immobile particles, with two particles belonging
to the same cluster if they are connected. The clusters are numbered
by $k=1,2,... K$ and we denote by ${\cal N}_k$ the k-th cluster with
$N_k$ particles. 
The clusters of mobile particles are defined in complete analogy.
However, in this case the identity of the clusters refers to a
particular snapshot --- usually the initial instant of time. In contrast
the clusters of immobile particles are likely to be time persistent
structures.  

\begin{figure}[tbp]
\includegraphics[width=3.2in]{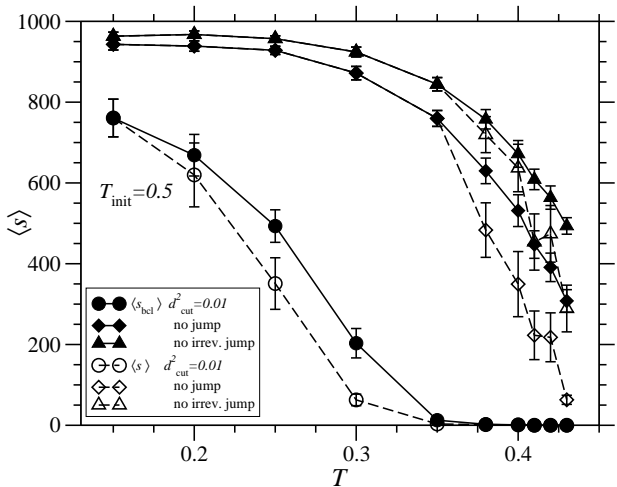}
\caption{\sf Average number of cluster members $s$ 
  as a function of temperature. Shown are an average of $s$ over only the 
  biggest cluster of each independent configuration 
  $\langle s_\ind{bcl} \rangle$ and an average over all clusters
  $\langle s \rangle$ are shown for definitions I, III 
  and IV.
}
\label{fig:sofT_dcut+nojump}
\end{figure}

To get a first idea of the organization of immobile particles into
clusters we simply determine 
\begin{equation}
\label{averagenumber}
\langle s \rangle=\langle \frac{1}{K}\sum_{k=1}^K N_k\rangle
\end{equation} 
i.e. the total number of immobile particles, $N_{immob}=\sum_{k=1}^K
N_k$, divided by the total number of clusters $K$, averaged over
initial conditions. Here and in the following $\langle \cdot \rangle$
denotes an average over initial conditions.  In
Fig.~\ref{fig:sofT_dcut+nojump} we show $\langle s \rangle$ (open
symbols) for the definitions I, III \& IV together with $\langle
s_\ind{bcl} \rangle$ (filled symbols), where
$s_\ind{bcl}=N_{k,\ind{bcl}}$ is the number of particles in the
largest cluster.  At the lowest temperatures almost all particles are
immobile (see Figs.~\ref{fig:QofT_dcut} and \ref{fig:QofT_nojump}) and
connected, so that there is only one large cluster and $\langle s
\rangle = \langle s_\ind{bcl} \rangle$ is approximately given by the
total number of particles.  As the temperature increases the fraction
of immobile particles decreases and the number of clusters decreases
simultaneously so that $\langle s_\ind{bcl} \rangle > \langle s
\rangle$. Definition IV (``no irrev. jump'') gives rise to a larger
$\langle s \rangle$ for immobile particles than definition III (``no
jump''), since IV excludes fewer particles from the subset than III,
namely only irreversible instead of also reversible jumpers.

To disentangle the effects of spatial rearrangements 
and changing number of immobile
particles, we use definition II with a {\em fixed} number of immobile
particles.
\begin{figure}[tbp]
\includegraphics[width=3.2in]{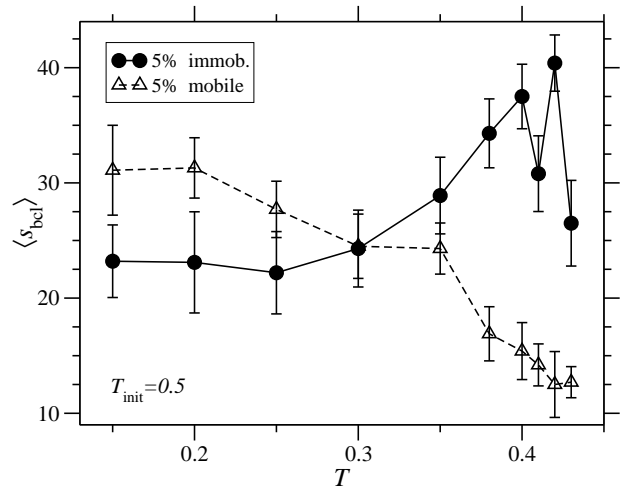}
\caption{\sf Same as Fig.~\ref{fig:sofT_dcut+nojump} but for definition 
   II and only for $\langle s_\ind{bcl} \rangle$.
}
\label{fig:sofT_perc}
\end{figure}
Fig.~\ref{fig:sofT_perc} shows $\langle s_\ind{bcl} \rangle$ for the 5\%
most mobile and 5\% most immobile particles as a function of
temperature. 
In the temperature range $0.25 \le T \le 0.4$
the 5\% immobile particles cluster increasingly more together
with increasing temperature. We interpret this such that immobile
particles manage to stay immobile at larger temperatures by forming each
others cages most successfully by clustering together. 
The 5\% most mobile particles on the other hand cluster more together at
lower temperatures. A more cooperative motion allows the particles to be
mobile at low temperatures.

\begin{figure}[tbp]
\includegraphics[width=3.2in]{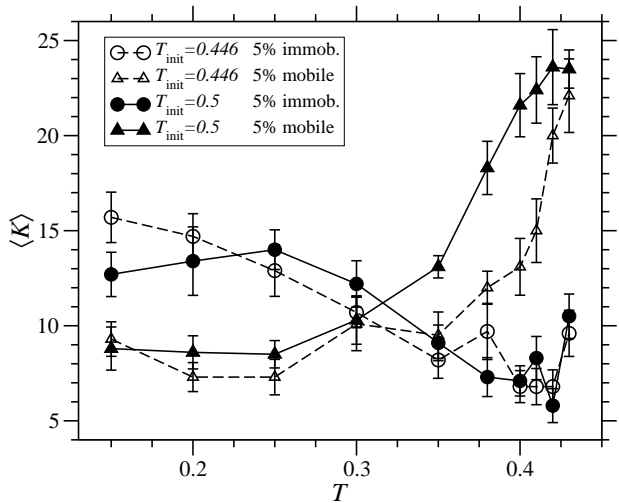}
\caption{\sf Average number of clusters  $\langle K \rangle$ as 
  a function of temperature using definition II.
}
\label{fig:KofT_perc}
\end{figure}
This picture gets further support from the observation that
 with increasing temperature the number of
clusters $\langle K \rangle$ increases for the 5\% most mobile
particles and
decreases for the 5\% most immobile ones (see
Fig.~\ref{fig:KofT_perc}). This temperature dependence is qualitatively
the same for simulations with different histories ($T_\ind{init}=0.446$
and $T_\ind{init}=0.5$). 

\begin{figure}[tbp]
\includegraphics[width=3.2in]{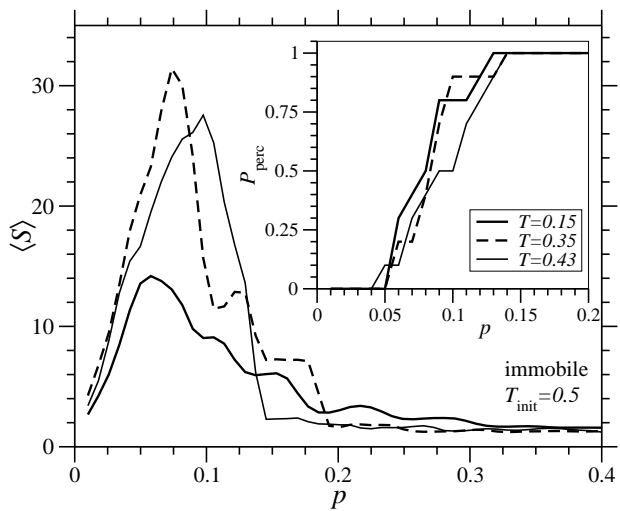}
\caption{\sf The average cluster size $\langle S \rangle$ as a 
  function of percentage $p$ using definition II for immobility.
  The average is over all non-percolating clusters and over 
  independent initial configurations. The inset shows the fraction of
  percolating configurations as a function of $p$.
}
\label{fig:Sofp}
\end{figure}

Another interesting question concerns the percolation properties of
clusters of either immobile or mobile particles.
We vary the percentage $p$ of immobile particles and 
thereby the clusters ${\cal N}_k$ as well as the total number of 
clusters $K$. For a fixed temperature $T$ we compute  the 
fraction $P_\ind{perc}$ of the ten independent configurations with a
percolating cluster of immobile particles. 
We define a cluster to be percolating if its
maximum extension $r_{\nu}=\mbox{max}
\, |{r_i}_{\nu} - {r_j}_{\nu}|$ between any pair of particles $i$ and
$j$ satisfies $r_{\nu} > 8.0$ for any direction $\nu\in \{x,y,z\}$
\cite{ref:unwrap}.  
As can be seen in the inset
of Fig.~\ref{fig:Sofp} this
fraction rises sharply from zero at a critical value $p_c$, which
is approximately independent of temperature.  
A central quantity from percolation theory is the average cluster
size, defined as 
\begin{equation}
\label{eq:S}
\langle S \rangle(p) =\left\langle \frac{\sum \limits_s s^2 n_s(p)}{\sum
  \limits_s s n_s(p)} \right\rangle
\quad \textrm{with} \quad
n_s=\sum_{k=1}^K \delta_{N_k,s}
\end{equation}
Percolation theory predicts a divergence of $\langle S(p) \rangle$, as the percolation
transition is approached from either side (excluding the percolating
cluster for $p>p_c$). In our case we observe a
strong increase of $\langle S \rangle$, a maximum around $p_c$
and a subsequent decrease for $p>p_c$ (see Fig.~\ref{fig:Sofp}). 
The strong increase of $\langle S \rangle$ around
$p_c$ is reminiscent of a percolation transition with, however, a
threshold value that is significantly lower than the corresponding
value for hard spheres \cite{ref:percolationHS}.  This may be due to
the strong correlations between immobile particles in
addition to the correlations due to the interaction potential $V(r)$.
Furthermore our analysis is severely hampered by the small size of our
sample, which percolates much easier than a larger sample, so that $p_c$
is underestimated.

\begin{figure}[tbp]
\includegraphics[width=3.2in]{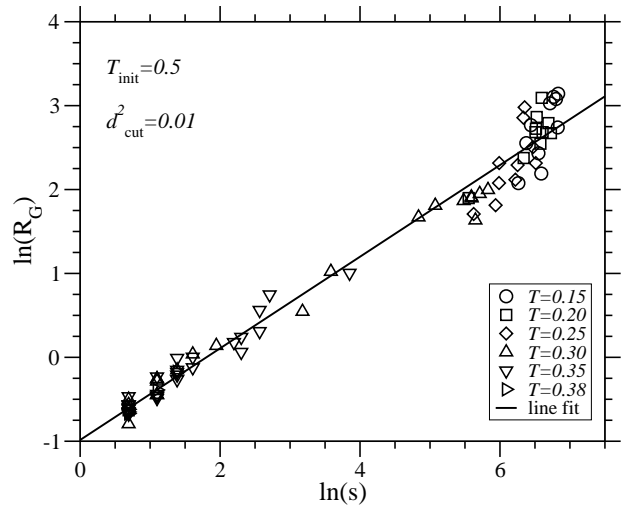}
\caption{\sf Log-log plot of the radius of gyration $R_\ind{G}$ 
  as a function of the number of cluster members $s$ using definition
  I for immobility. The line is 
  a linear fit with slope $0.55$.
}
\label{fig:lnRGoflns}
\end{figure}
The {\em spatial} extent of clusters of immobile and mobile
particles can be quantified with help of the radius of gyration. The
latter is defined for a particular cluster ${\cal N}_k$:
\begin{equation}
\label{eq:RG}
R_\ind{G}({\cal N}_k) = \sqrt{ \frac{1}{N_k} \sum_{i\in {\cal N}_k} 
     \left({\bf r}_i - {\bf r}_\ind{cm} \right)^2
                             }
\end{equation}
with the center of mass of the cluster
\begin{equation}
{\bf r}_\ind{cm}=\frac{1}{N_k} \sum_{i\in {\cal N}_k} {\bf r}_i
\end{equation}
To study the scaling of
$R_\ind{G}$ as a function of the number of particles in a cluster $s$,
we show in Fig.~\ref{fig:lnRGoflns} a log-log plot of
$R_\ind{G}({\cal N}_k)$ versus $s({\cal N}_k)$ for all clusters 
${\cal N}_k$ of all independent configurations.
We find that approximately $R_\ind{G} \propto s^{1/d_f}$ 
with a fractal dimension $d_f \approx 1.8$ 
(and similarly for the definitions II -- IV). If the
clusters were approximately straight lines one would expect
$d_f = 1.0$, if they were compact $d_f=3.0$. Our result points to
fractal structures, however
this should be taken with caveat since the clusters are too
small to observe true asymptotic scaling behavior.
\begin{figure}[tbp]
\includegraphics[width=3.2in]{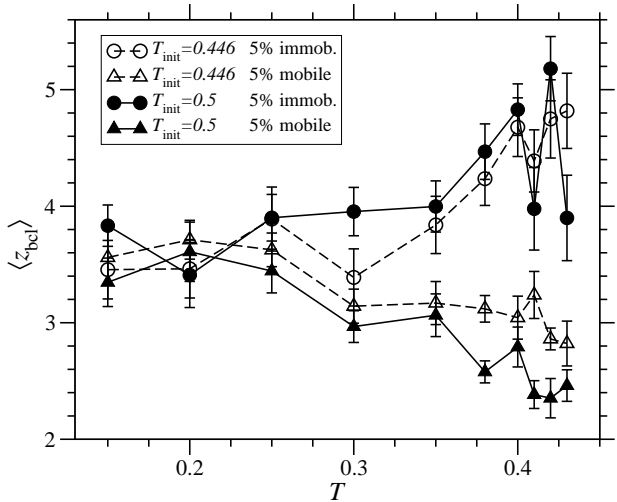}
\caption{\sf Average coordination number of members of the biggest 
  cluster $\langle z_\ind{bcl} \rangle$ as a function of temperature 
  using definition II.
}
\label{fig:zofT_perc}
\end{figure}

Next, we investigate the connectivity within single clusters. Since
two sites have been defined to be connected if sufficiently close in
space, the connectivity is not a purely topological feature, but 
gives us information about the shape of the clusters.
A convenient measure, which does
not require a large variety of cluster sizes, is the coordination
number $z$. 
We define the average coordination number $z({\cal N}_k)$ of a cluster
${\cal N}_k$ as
\begin{equation}
z({\cal N}_k) = \frac{1}{N_k} \, \sum \limits_{i \in {\cal N}_k} z_i
\end{equation}
where $z_i$ is the number of neighboring particles $j \in {\cal N}_k$
of particle $i$.  Fig.~\ref{fig:zofT_perc} gives a plot of $\langle
z_\ind{bcl} \rangle = \langle z({\cal N}_\ind{bcl}) \rangle$, the
average number of neighbors of particles in the biggest cluster
averaged over independent initial configurations.  We observe that
$\langle z_\ind{bcl} \rangle$ is increasing for the 5\% immobile
particles with increasing temperature, independent of the history of
the runs. The immobile particles block each other at larger
temperatures by clustering more together and by forming more compact
clusters. Clusters of mobile particles on the other hand become more
elongated with increasing temperature.  Although $\langle z_\ind{bcl}
\rangle$ is not equal to two as it would be for strings, $\langle
z_\ind{bcl} \rangle$ approaches 2.5 as $T \to T_g$. This is
reminiscent of the results
\cite{ref:mob_length+size,ref:PRE70,ref:eweeks}, where increasingly
string-like motion for mobile particles were observed as the glass
transition is approached from {\em above}.  (The authors of
\cite{ref:mob_length+size,ref:PRE70,ref:eweeks} use a slightly
different definition of mobility than in this paper and define a
string via velocity-velocity correlations.)

\begin{figure}[tbp]
\includegraphics[width=3.2in]{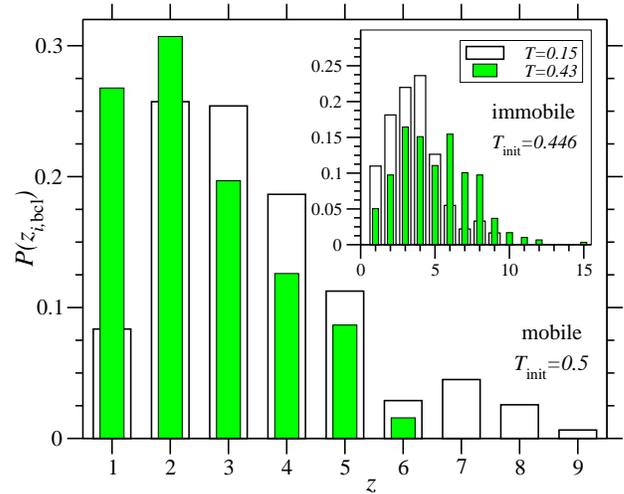}
\caption{\sf Distribution of coordination numbers $P(z_{i,\ind{bcl}})$ 
  of the 5\% fastest particles in the biggest cluster for
  simulations with $T_\ind{init}=0.5$.The inset shows $P(z_{i,\ind{bcl}})$ for 
  the 5\% slowest particles for simulations with $T_\ind{init}=0.446$.
}
\label{fig:Pofz}
\end{figure}

More information about the connectivity and the shape of the cluster 
can be found from the
distribution of coordination numbers  
of the particles in the biggest
cluster for each initial configuration
$P(z_{i,\ind{bcl}})= P(z_{i \in {\cal N}_\ind{bcl}})$.
Fig.~\ref{fig:Pofz} illustrates that for low temperatures $z=3$ and
$z=2$ are the most common coordination numbers of the mobile
particles in the biggest cluster. 
However, for high temperatures $z=2$
dominates and the variance of the distribution is smaller, 
indicating string-like clusters.
In the case of the 5\% immobile particles, on the other hand, the peak 
is at $z=4$ and the variance of $P(z_{i \in {\cal N}_\ind{bcl}})$ is
strongly increasing with increasing temperature (see inset of 
Fig.~\ref{fig:Pofz}). This indicates
increasingly more compact clusters with increasing temperature.

We finish our cluster analysis with ring statistics. Whereas ring
analysis usually is applied to network forming glasses, it is here
applied to the binary Lennard-Jones system, which is a fragile glass
former \cite{ref:fragileLJ}.
In the following we call any path a set of nearest neighbor connections.
The most commonly used method, introduced by S.\ V.
King \cite{ref:King}, searches for rings by finding for each particle
and any pair of its neighbors the shortest path between them.
As pointed out by Franzblau \cite{ref:Franzblau}, 
this definition has serious disadvantages 
such as not counting certain ``intuitive'' rings, which
leads Franzblau to the ``shortest-path'' (SP) criterion:
an SP ring contains for each pair of particles in the ring the shortest 
path between the pair in the cluster under consideration.
We use the SP criterion and search algorithm of Franzblau to 
find within the largest cluster all rings $r=1,N_{r,\ind{bcl}}$ each 
of size $n_r$, i.e.\ with $n_r$ connections.

\begin{figure}[tbp]
\includegraphics[width=3.2in]{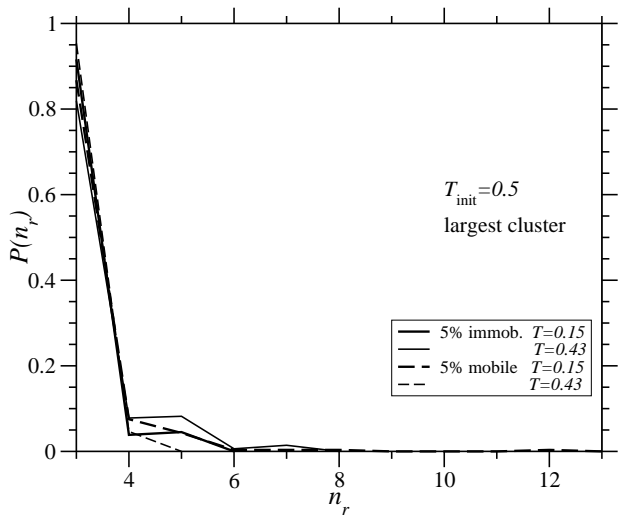}
\caption{\sf Distribution of ring sizes $P(n_r)$ within the 
   largest cluster of each independent configuration using definition II.
}
\label{fig:Pofn_perc}
\end{figure}

Fig.~\ref{fig:Pofn_perc} shows the distribution $P(n_r)$ for the biggest 
cluster of the 5\% immobile and 5\% mobile 
particles. We find that  rings are dominantly made 
up of three particles, only 
a few rings are longer. This means that in the case of the ``stringier''
mobile particle clusters, the clusters are mostly not closed rings. 

\begin{figure}[tbp]
\includegraphics[width=3.2in]{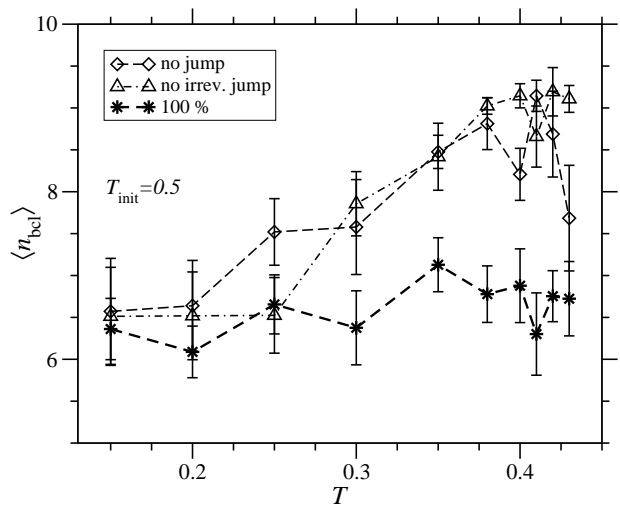}
\caption{\sf Average ring size $\langle n_\ind{bcl} \rangle$ as 
  a function of temperature using definitions III and IV and the 
  comparison of $\langle n \rangle$ for the complete system (100\%).
}
\label{fig:nofT}
\end{figure}

In the case of immobile clusters using definitions I, III and IV larger
clusters are formed leading to larger average ringsize
\begin{equation}
\langle n_\ind{bcl} \rangle = \langle \frac{1}{N_{r,\ind{bcl}}} \, 
   \sum \limits_{r=1}^{N_{r,\ind{bcl}}} \, n_r \rangle
\end{equation}
(see Fig.~\ref{fig:nofT}). For low temperatures almost all particles are
members of the biggest cluster of immobile particles and $\langle n_\ind{bcl}
\rangle$ is therefore comparable to $\langle n \rangle$ of the
whole system. For larger temperatures the clusters become more compact
and on average larger ring sizes occur.

\section{Conclusions}
\label{sec:conclusions}

We investigate a binary Lennard-Jones system below the glass transition.
To study the inhomogeneous structure of the glassy state
quantitatively we focus on the
most mobile and immobile particles. Since the identification of these
is not unique --- at least not in a simulation run of finite length ---
we test different definitions of
immobility. The first two definitions are based on the vibrational
amplitude and the last two define a particle to be immobile if it is not
jumping. 

The length scales of single particle motion varies strongly among the
particles, giving rise to a broad distribution of
 vibrational amplitudes for mobile and immobile particles.
The range of vibrational amplitudes increases as the glass transition
temperature is approached from below: The system does not only become
more heterogeneous with increasing temperature, but also the range of 
lengthscales extends to much larger scales in
the vicinity of the glass transition than at low temperatures. 
Focussing on the immobile particles only, we observe the distribution of
vibrational amplitudes to follow a scaling plot close to the glass transition.
It even compares reasonably well with a 
mean field theory for the amorphous solid state, developed in the
context of the gelation transition. Tentatively this may be taken as
evidence for the recently suggested picture of the more general
jamming transition \cite{ref:LiuNagel}, including both
the gel transition and the glass transition.

To characterize the spatial arrangement and correlations among
immobile and mobile particles, we identify clusters of particles,
which are defined to be connected if their relative distance is
smaller than a threshold value, given by the first minimum of the pair
correlation function. Whereas for the immobile particles these clusters
are approximately time persistent and hence characterize the static
heterogeneity, the clusters of mobile particles are relevant for the
dynamic heterogeneity of the glassy state. 

If immobility is defined by the vibrational amplitudes or the jump
characteristics, then the number of immobile particles increases
strongly as the temperature is lowered. This effect dominates the
statistics of cluster number, size and shape.
Baljon {\it et al.\ 
}\cite{ref:Baljon} find similar results for thin polymer films by
looking at the percolation probability of immobile monomers as a
function of temperature. 
It is instructive to keep the number of immobile/mobile particles constant, namely the $p$\%
particles with the smallest/largest vibrational amplitude. 
The $p$\% mobile particles on the other hand cluster more together at
low temperatures and form more elongated string-like clusters at
larger temperatures.  Similar strings have been found for the same
system above the glass transition \cite{ref:Donati1998,ref:PRE60} and
for a similar system above and slightly below the critical temperature
\cite{ref:JCP120}.
These simulations, our results as well as experiments
\cite{ref:eweeks} point to string-like clusters of mobile particles 
which grow close to the glass transition temperature, 
both when approached from 
above as well as when approached from below the glass transition.
We find
that with increasing temperature the $p$\% immobile particles cluster
more strongly together and form more compact clusters which allows the
particles to block each other more effectively, and to therefore stay
immobile.

\section{Acknowledgments}

KVL gratefully acknowledges financial support from DFG Grant No. Zi 209/6-1.
We thank W. Kob for the input in the early stages of this work and 
B. Vollmayr-Lee for the careful reading of the manuscript.



\begin{thebibliography}{}
\bibitem{ref:errorbars} Here and for all other figures errorbars are
  determined solely via the average over the ten independent
  configurations.
\bibitem{ref:unwrap} For the definition of $r_{\nu}$ as well as for 
  $R_\ind{G}$ (see below) we have to make sure to
 eliminate the effect of periodic boundary conditions. We therefore first
 unwrap each cluster
 i.e.\ periodic images are chosen such
 that all neighboring particles satisfy directly $r_{ij} \le r_\ind{min}$.
\end{thebibliography}
\end{document}